\documentclass[epj,referee]{svjour}

\usepackage{graphics}
\usepackage{color}

\begin{document}
\title{The actin cortex as an active wetting layer}
\author{J.-F. Joanny\inst{1} 
\and K. Kruse\inst{2} 
\and J. Prost\inst{1,3}
\and S. Ramaswamy\inst{4}\thanks{\emph{on leave from Department of Physics, Indian 
Institute ofScience, Bangalore 560 012}}}

\institute{Physico Chimie Curie (Institut Curie, Cnrs UMR 168, UPMC), 
Institut Curie Centre de Recherche, 26 rue d'Ulm 75248 Paris Cedex 05, France 
\and Theoretische Physik, Universit\"at des Saarlandes, Postfach 151150,
66041 Saarbr\"ucken, Germany
\and ESPCI, 10 rue Vauquelin 75005 Paris, France
\and TIFR Centre for Interdisciplinary Research, 21 Brundavan Complex,
Narsingi, Hyderabad 500 075 India}

\date{Received: \today}

\abstract{
Using active gel theory we study theoretically the  properties of the cortical actin layer of animal cells. The cortical layer is
described as a non-equilibrium wetting film on the cell membrane. The actin density is approximately constant in the layer
and jumps to zero at its edge. The layer thickness is determined by the ratio of the polymerization velocity and the depolymerization
rate of actin.}

\maketitle

\section{Introduction}

Living cells maintain and change shape, adhere, spread, divide and
crawl through the agency of a dynamic, filamentous scaffold known as the
cytoskeleton~\cite{albe08}. Although this structure is made up of many different
constituents the task of stress generation, with crucial consequences for the
mechanical properties of cells, lies primarily with the \textit{acto-myosin}
component~\cite{howa01}. This substructure consists of a meshwork of semi-flexible actin
filaments interacting with a large number of proteins among which myosin
molecular motors play a major role. Myosin motors, assembled in 
minifilaments, consume free energy through the hydrolysis of ATP molecules and
can produce work. By binding to the actin filaments, myosin minifilaments create
contractile stresses in the actin gel. 

The crosslinked polymer network formed by
actin and its associated proteins differs profoundly from more familiar
thermal-equilibrium physical or chemical gels because of the sustained energy
dissipation by the molecular motors and some other proteins. It is usefully
viewed as a state of active matter \cite{rmp} known as an active polar gel
\cite{krus04,krus05,joan07,hatw2004,juli07} -- ``active'' referring to the steady consumption
of free energy at the scale of individual components, i.e., the actin-bound
molecular motors, and ``polar'' to the orientable, directed character of the actin
filaments. In particular, each filament carries a distinction between its two
extremities, as does each constituent monomer. We will therefore refer to the
plus and the minus end of a filament.

The filaments of the cytoskeleton undergo constant assembly and disassembly:
Each actin filament polymerizes preferentially at the plus end and depolymerizes
at the minus end. This process is called treadmilling, and is regulated by
a multitude of accessory proteins. Through depolymerization, filaments
disintegrate, and through polymerization new filaments are generated,
maintaining a nonzero gel mass on average. Generation of new filaments is
assisted by nucleating proteins, such as formins that act as seeds for
actin polymerization and then stay attached for some time to actin plus-ends,
where they promote the addition of monomers. The Arp2/3 complex instead binds to
existing filaments and remains attached to the minus end of newly
created actin filaments.

In this work we study consequences of the interplay of actin polymerization
and active contraction of actin gels through a hydrodynamic description,
that captures the generic behavior of materials on large length and time scales.
Such a description starts with the correct choice of slow variables~\cite{mart72}
relying  broadly on conservation laws, broken continuous symmetries, and
order-parameter modes near a continuous phase transition. This approach has
been successfully extended to active systems~\cite{rmp,krus04,tone98,simh02}, held
in stationary states far from thermal equilibrium by the sustained dissipation
of free energy. Despite the very broad range of length scales, from microns to
kilometers, on which organized active matter is seen, there is a degree of
universality in its hydrodynamic properties, classified by the type of broken
symmetry and applicable conservation laws: examples include waves without
conventional inertia, anomalously large number fluctuations, and a tendency of
instability of quiescent states towards spontaneous flow~\cite{rmp}.

\begin{figure}
\resizebox{0.5\textwidth}{!}{
 \includegraphics{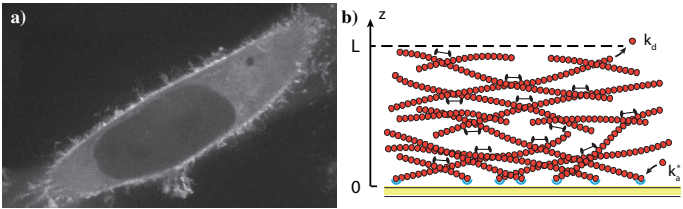}
 }
   \caption{a) Confocal fluorescence microscopy image of a HeLa cell with F-actin labeled by life-act ruby. Below the 
   plasma membrane a well-defined zone of high fluorescence represents the actin cortex. Image with courtesy from 
   M. Fritzsche and G. Charras.
   b) Schematic illustration of the cortex dynamics considered in this work. Nucleation promoting factors located at the
   plasma membrane nucleate new actin filaments or assist elongation of existing filaments at rate $k_a^*$. This generates
  a flux of polymerized actin $v_p\rho_0$, where $\rho_0$ is the gel density at the surface and 
   $v_p=k_a^*\delta$  is the polymerization velocity,
$\delta$ being the size of an actin monomer.
Filaments disassemble at 
   a rate $k_d$ anywhere in the cortex. Not shown is the growth of actin filaments away from the membrane. Molecular
   motors act as active cross-links and generate active stresses in the cortex.
   }
 \label{fig:sketch}
\end{figure}

We focus on the fact that a substantial fraction of the
actomyosin is located in a layer, known as the actin cortex~\cite{salb07,meda02}, adjacent to the
plasma membrane of the cell, see Fig.~\ref{fig:sketch}a. This localization of one component to the vicinity
of a wall is reminiscent of the physical phenomenon of wetting.
However, prevalent explanations of the cortical actin profile do not take
advantage of this analogy, relying instead on an imposed spatial separation
between the zones of polymerization and depolymerization. In this paper we show
that the active contractility alluded to in an earlier paragraph can drive a
transition to a steady state maintained by a polymerization-depolymerization
process, see Fig.~\ref{fig:sketch}b, in which the polymer concentration has a profile very similar to that
of a wetting layer~\cite{cahn77}. The cortex thickness and actin density profile emerge
naturally from our treatment. Recall that in thermal equilibrium systems wetting
arises through the selective attraction of a component to a surface. In the
following we show that contractility, although quite different from an
attractive potential, plays a similar role in driving a condensation at the
plasma membrane. To obtain this result we generalize the hydrodynamic
description of active gels to include crucial nonlinear density dependences.

The rest of this paper is organized as follows. In section II we present
the hydrodynamic equations for polymerizing active gels. 
In Section III, we study an active gel polymerizing at a surface, which yields a
description of the actin cortex of animal cells. The properties of the gel can
be understood in terms of active wetting and dewetting by filament assembly and
disassembly.

\section{Hydrodynamic description}

We now present the hydrodynamic equations governing the dynamics of an assembling 
active actin gel. We assume
that the gel is assembled by actin polymerization at a surface.  In a cell or in 
an in vitro experiment, the polymerization is promoted by nucleating proteins, 
such as proteins of the formin family, see Fig.~\ref{fig:sketch}b. Away from the 
surface, the actin gel assembles by elongation of existing 
filaments or by nucleation of new filaments. The gel 
disassembles because of monomer 
removal at filament minus-ends or by severing of the gel filaments that produces 
small filaments, which diffuse into the solution. Myosin molecular motors assemble 
into small filaments that act as cross-links, which actively generate mechanical 
stress in the 
filament network. The network is permeated by a solvent containing, in particular, unbound 
motors and actin monomers. 
We limit ourselves to the case where the exchange of motors between the actin 
network and the solvent is so 
fast that we can assume them to be equilibrated and where the motors diffusion is 
fast enough that the concentration of free motors in the solvent is constant. In a 
first approximation, the concentration of motors bound to actin is then proportional 
to the 
local actin concentration. We have verified numerically that our results do not 
change significantly when motor binding and unbinding and motor diffusion are 
taken 
explicitly into account, at least in the case when the gel does not impede the diffusion of motors. 

As mentioned in the introduction, actin filaments are polar objects. Consequently, 
an actin gel can present a macroscopic 
vectorial order represented by a polarization field $\mathbf{p}$. 
In the present treatment we ignore the subtleties associated with actin 
polarization and consider in the following the case of an isotropic gel.
Our detailed calculations will be for the case where the actin concentration
varies only along one direction, where transverse variations in
$\mathbf{p}$ play no role.

Actin polymerizes on a planar surface located in the 
plane $z=0$. The gel and the solvent are restricted to the half-space $z\ge0$. If 
the gel is homogeneous in the directions parallel to the surface, its 
properties and in particular its density only depend on the $z$ coordinate.
We will explore elsewhere the dynamics of variations in the $xy$-plane.

\subsection{Conservation laws}

The hydrodynamic description is based on conservation laws for mass and momentum. 
Mass conservation of the gel 
and the solvent read
\begin{eqnarray}
\partial_t\rho_\mathrm{g} +\partial_\alpha\rho_\mathrm{g} v_{\mathrm{g},\alpha} &= &
-k_d\rho_\mathrm{g}\nonumber\\
\partial_t\rho_\mathrm{s} +\partial_\alpha\rho_\mathrm{s} v_{\mathrm{s},\alpha} 
&= &k_d\rho_\mathrm{g}\quad.
\end{eqnarray}
Greek indices denote the three spatial directions $x$, $y$, and $z$ and we have 
adopted Einstein's summation convention.

In the above expressions, $\rho_\mathrm{g}$ and $\rho_\mathrm{s}$, respectively, 
denote the densities of the gel and the 
solvent, while $\mathbf{v}_\mathrm{g}$ and $\mathbf{v}_\mathrm{s}$ are the 
corresponding velocities. Degradation 
of the gel due to depolymerization and severing occurs at a constant rate $k_d$. The 
gel is produced on the one hand by growth of existing filaments, and on the other 
hand by the nucleation of new actin filaments. There is no significant spontaneous 
nucleation of new filaments at concentrations 
of monomeric actin present in cells. Instead nucleation promoting factors  
regulate the generation of new filaments and their elongation. 
Members of the formin family and the Arp2/3 complex are important examples of 
nucleation promoting factors~\cite{rome04,bugy10,acha10}. Formin is located directly 
beneath the plasma membrane, where it generates new filaments. The Arp2/3 complex 
needs to bind to existing filaments 
before it can act as the seed of a new filament. Still, it is predominantly 
localized in the vicinity of the cell membrane. We therefore consider here for 
simplicity that the actin gel growth occurs only at the surface. Therefore, the filament density far away from
the surface vanishes. We account for 
elongation and nucleation 
on the surface by a boundary condition on the gel flux 
at $z=0$. Explicitly, we write 
$\left.\rho_\mathrm{g} v_{\mathrm{g},z}\right|_{z=0}=v_p\rho_0$, where $v_p$ is 
the polymerization velocity and $\rho_0$ 
the gel density at the surface, which is imposed by the density of 
nucleation promoting factors. 

\subsection{Constitutive relations}

To fully specify the behavior of the actin gel, we must provide constitutive 
equations. They link the generalized thermodynamic fluxes
to the generalized thermodynamic forces. We follow here closely the approach of 
Callan-Jones and J\"ulicher~\cite{call11} to describe active permeating gels. The 
fluxes are in our case the gel 
velocity,  the time derivative of the strain in the gel, the relative 
current between actin gel and solvent $\mathbf{j}=\rho_\mathrm{g} \left( 
\mathbf{v}_\mathrm{g} - \mathbf{v} \right)$, where 
$\mathbf{v}$ is the center of mass velocity, the deviatoric stress 
tensor $\mathsf{\sigma}$ and the rate of ATP consumption. The generalized forces
are the gradient in the relative chemical potential $\bar\mu$ of the gel and the 
solvent, the center of mass velocity gradient $\nabla \mathbf{v}$, 
the partial stress of the gel $\mathsf{\sigma}^\mathrm{g}$ and the activity of the 
system. The relative chemical potential is given by
$\bar\mu=\mu_\mathrm{g}/m_\mathrm{g}-\mu_\mathrm{s}/m_\mathrm{s}$, where 
$\mu_\mathrm{g}$ and 
$\mu_\mathrm{s}$ are the respective chemical potentials of the gel and the solvent 
and $m_g$ and $m_s$ the masses of the solvent and actin monomer molecules. Active 
processes are eventually driven 
by the hydrolysis of ATP into ADP and inorganic phosphate P$_i$, with
chemical potentials $\mu_\mathrm{ATP}$, $\mu_\mathrm{ADP}$, and
$\mu_\mathrm{P}$, respectively. We express the system's activity through the difference
$\Delta\mu=\mu_\mathrm{ATP}-\mu_\mathrm{ADP}-\mu_\mathrm{P}$. In the 
following $\Delta\mu$ is considered as constant in space and
time. We do not consider any further the corresponding thermodynamic flux which 
provides the rate $r$ of ATP consumption in the system.

The constitutive equations are 
obtained from an expansion of the thermodynamic fluxes in terms of the 
corresponding forces. Here, we give only the final 
equation for the relative current between actin and solvent for an isotropic 
active gel \cite{call11}
\begin{equation}
j_\alpha= 
-\gamma\partial_\alpha\bar\mu+\chi\partial_\beta\sigma^\mathrm{g}_{\alpha\beta} \quad.
\label{eq:j}
\end{equation}
In this expression $\mathsf{\sigma}^\mathrm{g}_{\alpha\beta}$ is are the components of the partial stress tensor of 
the gel phase, not to be confused with the total stress~\cite{call11}. It is analogous to the particle-phase stress in 
suspension mechanics~\cite{batc70,nott94}
In steady state elastic effects 
vanish and the stress is purely viscous, so that
$\sigma^\mathrm{g}_{\alpha\beta}= 2\eta v_{\mathrm{g}, 
\alpha\beta}- \zeta\Delta\mu\delta_{\alpha\beta}$ 
where $\eta$ is the gel viscosity and $v_{\mathrm{g},
\alpha\beta}= \frac{1}{2} \left( \partial_\alpha v_{\mathrm{g},\beta}+
\partial_\beta v_{\mathrm{g},\alpha}\right)$ and $\zeta\Delta\mu$ gives the magnitude of
the actively generated (isotropic) stress in the gel phase. 
The first term in expression (\ref{eq:j}) accounts for the diffusive current 
due to gradients in the relative chemical potential.  The second term describes 
the flux that results from gradients in the 
gel stress. Finally, the permeation constant $\rho_\mathrm{g}\chi^{-1}\sim\eta_s/\xi^2$, where $\xi$ is 
the gel's mesh size and $\eta_s$ the solvent viscosity. 

 As the 
only vector for a non polar system is the gradient vector, 
within the Onsager linear approach the active contribution to the current is 
proportional to $\partial _\alpha \Delta \mu$ and vanishes if $\Delta \mu$ is 
constant in space, which we assume. 
Fluxes stemming from the system's activity are captured by the last term 
of the stress expression.
The coupling parameter $\zeta$ here depends  
on the gel density $\rho_\mathrm{g}$ and is negative due to the 
contractility of the motors.

We proceed using the same approximation as in Ref.~\cite{call11}: 
The osmotic pressure $\tilde \Pi$ satisfies the 
Gibbs-Duhem equation $d {\tilde \Pi}=\rho_\mathrm{g}d\mu$, 
if the volume of the system is independent of
the composition and incompressible. We define an effective osmotic pressure 
including active effects as $\Pi={\tilde \Pi} +\zeta\Delta\mu$ and, we thus get
\begin{equation}
\label{perm}
\frac{\rho_\mathrm{g}}{\chi}\left(v_{\mathrm{g},\alpha} - v_\alpha\right)
=2\eta\partial_\beta v_{\mathrm{g},\alpha\beta}-\partial_\alpha \Pi,
\end{equation}
where we have expressed the gel stress in terms of the gel shear 
rate. There are two competing dissipative mechanisms, the gel viscosity and 
permeation of the solvent through the actin gel. The comparison between these two 
types of 
dissipation defines a 
permeation length $L_p=(\eta \rho_\mathrm{g}^{-1}\chi)^{1/2 } \sim (\eta/\eta_s)^{1/2}\xi$. 
The relative permeation current  
on the left hand side of Eq.~(\ref{perm})
is negligible compared to the viscous dissipation term if we consider 
the dynamics on length scales smaller than the permeation
length scale $L_p$. For typical viscosities $\eta\sim10^{8}\eta_s$ 
and typical mesh sizes of $\xi$ of a few tens of nanometers, the 
characteristic length is of the order of a few hundreds of microns and thus macroscopic. We therefore neglect 
permeation in the following. 

From now on, we consider only the gel density $\rho_\mathrm{g}$ and the 
corresponding velocity field $v_\mathrm{g}$.
Dropping the $g$ indices, the equations read
\begin{eqnarray}
\nonumber
\partial_t\rho+\partial_\alpha\rho v_\alpha &=& - k_d \rho\\
\label{eq:stress}
2\eta\partial_\beta v_{\alpha\beta} - \partial_\alpha\Pi(\rho) &=&0.
\end{eqnarray}
These equations are identical to those of a hydrodynamic theory for an effective one component compressible active 
gel~\cite{krus04,krus05}.  

\section{Active pre-wetting}

We consider a situation where the gel is assembling at the surface located at $z=0$ and assume invariance under 
translations in the surface plane. This leaves us with a one-dimensional problem.
The dynamic equations (\ref{eq:stress}) read for $z\ge0$:
\begin{eqnarray}
\nonumber
\partial_t\rho+\partial_z\rho v &=& - k_d \rho\\
\label{eq:stress1d}
\eta\partial_z v -  \Pi(\rho) &=&0,
\end{eqnarray} 
where $v$ denotes the $z$-component of the gel velocity. The second equation follows from integrating 
Eq.~(\ref{eq:stress}). Let us recall that the boundary condition on the gel current 
at the surface is  $\rho v|_{z=0}=\rho_0 v_p$. 

To complete our description, we need to provide an expression for the effective 
pressure $\Pi$. 
Usually, the osmotic pressure $\tilde \Pi$ is a monotonic increasing function of 
the actin density. However, the active contribution is negative and  has a maximum as a function
of density for a given crosslink density~\cite{bend08}.  
The passive osmotic pressure always dominates at large densities but if activity  is large 
enough, the effective osmotic pressure can become a non monotonic function 
of density, which is indicative of a phase separation in the solution induced by the 
contractility of the molecular motors.
In the following, we use
\begin{equation}
\Pi = a\rho^3+b\rho^4\quad,
\end{equation}
where the coefficient $a$ depends on activity.
For vanishing activity, standard three body interactions should dominate
and $a$ should be positive. For large activity, contractility should dominate and $a$ should be negative. In principle,
the expansion in powers of $\rho$ should contain linear and quadratic terms as well. Whether contractility should be reflected
in the quadratic or in the cubic term depends on how the motor density relates
to the actin polymeric density. For example, one could argue that
contractility requires a pair of actin filaments as well as myosin, so if the
bound myosin concentration is proportional to that of the filamentous actin,
one would expect a cubic $\rho$ dependence. The choice of
the functional form is not essential, what really matters is the non monotonicity of $\Pi(\rho)$. 
If $a<0$ then the activity leads to contractile stresses.

We solve Eq.~(\ref{eq:stress1d}) in steady state. Combining the two equations, we get for the velocity field 
\begin{equation}
\label{eq:vsteadystate}
\eta v=\left(\frac{\partial\rho}{\partial z}\right)^{-1}\rho f(\rho)\quad,
\end{equation}
where we have introduced the auxiliary function 
\begin{equation}
f(\rho)=-k_d\eta-a\rho^3-b\rho^4.
\end{equation}
The general solution to the steady state equations can then be written in the form
\begin{equation}
\label{eq:drhodzsteadystate}
\frac{d\rho}{dz}=\frac{1}{\rho_0\eta v_p}\rho^2 f(\rho)\exp\left\{\int_{\rho_0}^\rho\frac{k_d\eta}{\rho'f(\rho')}d\rho'\right\}
\end{equation}
leading to
\begin{equation}
\label{eq:vsteadystateexplicit}
v=\frac{\rho_0}{\rho}v_p\exp\left\{-\int_{\rho_0}^\rho\frac{k_d\eta}{\rho'f(\rho')}d\rho'\right\}\quad.
\end{equation}

From these solutions we infer that the gel density $\rho$ jumps to zero at densities fulfilling $f(\rho)=0$. 
Indeed, equation~(\ref{eq:vsteadystateexplicit}) implies $v\ge0$ with $v=0$ for $f=0$ and no material is 
transported beyond such a point. Consequently, as soon
as $f$ presents zeros a well-defined cortical layer with a sharp edge can be formed. The existence of zeros
of $f$ depends on the activity. There exists a critical value $a_c$ beyond which a wetting layer appears. 
For $-a>-a_c>0$ the function has two zeros, while for 
$-a<-a_c$ it has none. If the activity is equal to the critical value $a_c$, the function $f$
has exactly one zero. As we will discuss below, the transition at $a=a_c$ can be viewed as an
active or non-equilibrium analog of a pre-wetting transition~\cite{cahn77,brez83}.

\subsection{The weak contractile activity regime}

The weak contractile activity regime is defined by $-a<-a_c$, such that $f(\rho)<0$ for all $\rho$. Since 
$\rho$ tends to zero as $z\to\infty$, we have in this case $f(\rho)\approx-k_d\eta$ and thus
\begin{equation}
\frac{d\rho}{dz}\approx-\frac{k_d}{\rho_0v_p}\rho^2\exp\left\{-\ln\frac{\rho}{\rho'_0}\right\}.
\end{equation}
where $\rho'_0$ is a constant. The density thus eventually decays exponentially. If $\rho_0$ is small, we have 
$\rho'_0\approx\rho_0$ and
\begin{equation}
\frac{d\rho}{dz}= -\frac{k_d}{v_p}\rho\quad,
\end{equation}
such that the characteristic length is $v_p/k_d$.
For large values of $\rho_0$, the characteristic length is instead $v_p\rho_0/k_d\rho'_0$ with 
\begin{equation}
\rho'_0=\rho_0\exp\left\{\int_0^{\rho_0}\frac{k_d\eta}{\rho'}\left(\frac{1}{f(\rho')}+\frac{1}{k_d\eta}\right)d\rho'\right\}.
\end{equation}

The profile can thus be a simple exponential or it can present a "shoulder" presaging the
existence of a well-defined cortical layer for high contractile activity. In the latter case, the profile presents two inflection
points, which are determined by $\partial (\partial_z\rho)/\partial\rho=\infty$. This equation has
two solutions for $-a$ large enough but still smaller than $-a_c$. In this case, we can
define the layer thickness $L$ via the condition of mass conservation in steady state 
$\rho_0v_p\sim k_d\rho_mL$, where $\rho_m$ is 
approximately given by the value of $\rho$ maximizing $f$. 

\begin{figure}
\resizebox{0.4\textwidth}{!}{
  \includegraphics{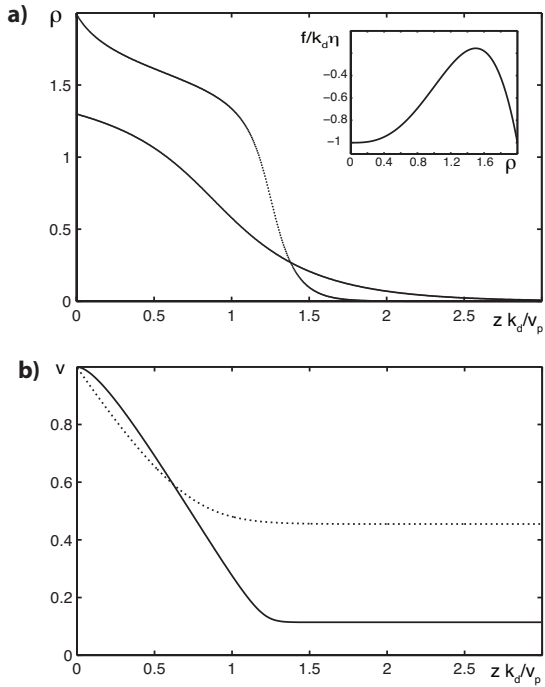}
}
   \caption{Steady state density and velocity profiles in the weakly active regime, $-a<-a_c$ from
   numerical solutions of Eqs.~(\ref{eq:stress1d}).
   a) Densities as a function of the distance from the polymerizing surface for surface densities $\rho_0=1.3$ and $\rho_0=2.0$,
   respectively. Inset: Auxiliary function $f(\rho)$. b) Corresponding velocity profiles; dotted: $\rho_0=1.3$ and solid: $\rho_0=2.0$. 
   Parameters are 
   $a=-k_d\eta$ and $b=0.5 k_d\eta$.}
 \label{fig:lowactivity}
\end{figure}

Examples of density profiles for a subcritical activity and two different values of $\rho_0$ are given in Fig.~\ref{fig:lowactivity}.

\subsection{The large contractile activity regime}

We consider now the case $-a>-a_c$, such that $f(\rho)$ has two real roots $\rho_1>\rho_2$.
From Equation~(\ref{eq:vsteadystate}) and since $v\ge0$ we infer that for a density $\rho(z=0)=\rho_0<\rho_2$ the density
approaches 0 with increasing $z$. Indeed, in this case $\partial\rho/\partial z<0$ for all $z>0$. As a consequence, the 
density profile behaves very similarly to the weak activity case $-a < -a_c$. In contrast, for $\rho_0>\rho_2$, the
density approaches the value $\rho_1$. As $v\to0$ for $\rho\to\rho_1$, this value is reached at a finite distance
$z$ from the surface.

\begin{figure}
\resizebox{0.4\textwidth}{!}{
  \includegraphics{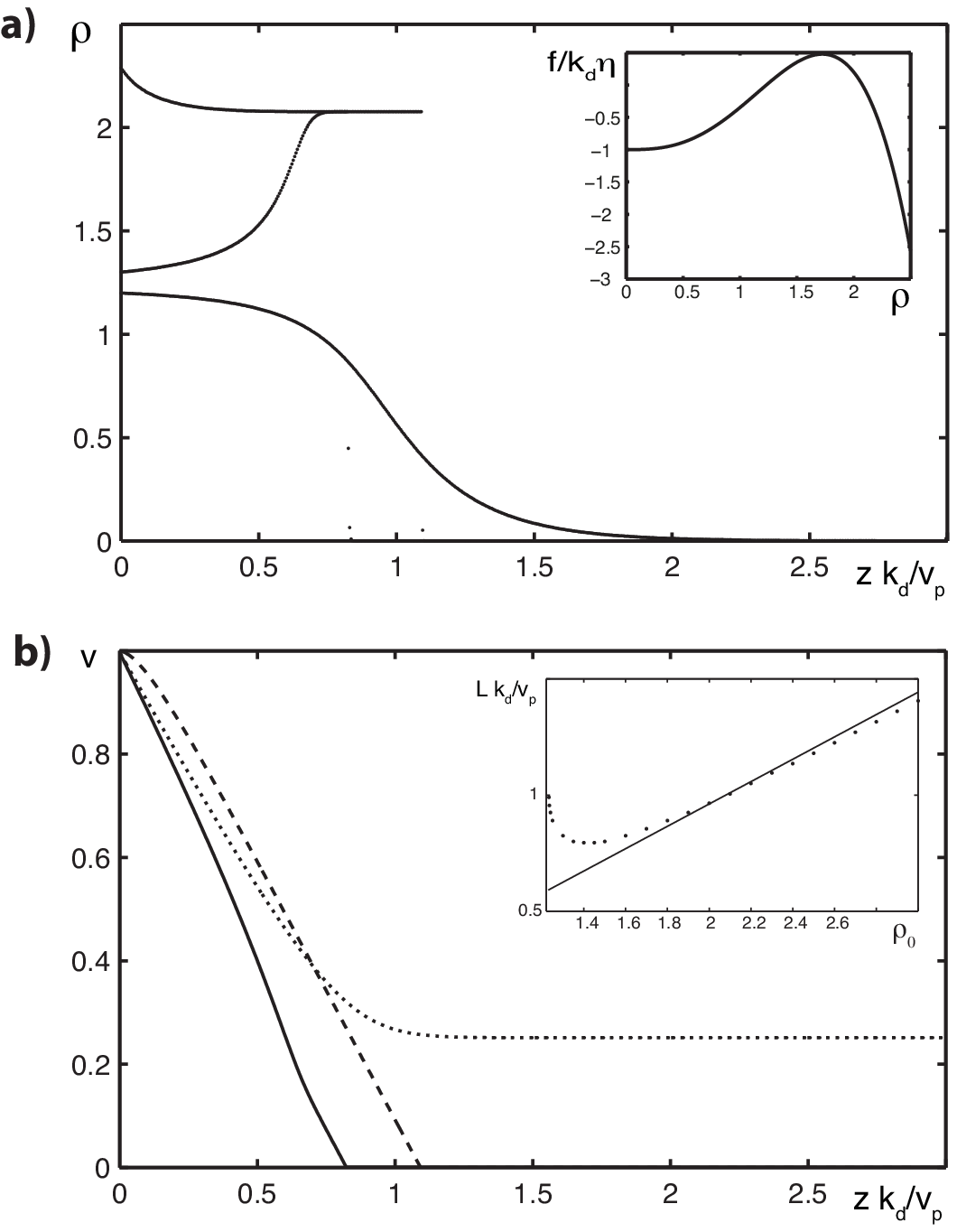}
}
   \caption{Steady state density and velocity profiles in the strongly active regime, $-a>-a_c$, from
   numerical solutions of Eqs.~(\ref{eq:stress1d}).
   a) Densities as a function of the distance from the polymerizing surface for surface densities 
   $\rho_0=1.2$, $\rho_0=1.3$, and $\rho_0=2.3$,
   respectively. Inset: Auxiliary function $f(\rho)$, which has zeros at $\rho_1=2.08$ and $\rho_2=1.23$. 
   b) Corresponding velocity profiles; dotted: $\rho_0=1.2$, solid: $\rho_0=1.3$, and dashed: $\rho_0=2.3$.
   Inset: cortex width as a function of $\rho_0$ from
   numerical solution of Eqs.~(\ref{eq:stress1d})  (dots) and from Eq.~(\ref{eq:L}) (line). Parameters are 
   $a=-1.15 k_d\eta$ and $b=0.5 k_d\eta$.}
 \label{fig:highactivity}
\end{figure}

\begin{figure}
\resizebox{0.4\textwidth}{!}{
  \includegraphics{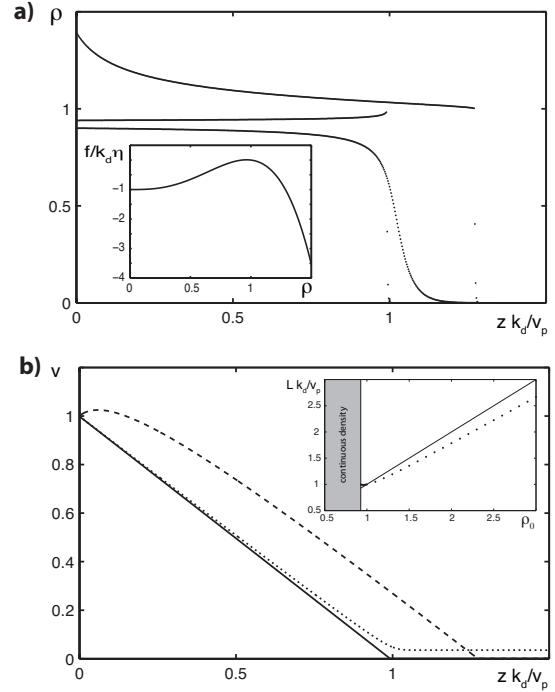}
}
   \caption{Steady state density and velocity profiles in the strongly active regime, $-a>-a_c$, from
   numerical solutions of Eqs.~(\ref{eq:stress1d}).
   a) Densities as a function of the distance from the polymerizing surface for surface densities 
   $\rho_0=0.9$, $\rho_0=0.94$, and $\rho_0=1.4$, respectively. Inset: Auxiliary function $f(\rho)$, 
   which has zeros at $\rho_1=1.0$ and $\rho_2=0.93$. b) Corresponding velocity profiles; dotted: $\rho_0=0.9$, 
   solid: $\rho_0=0.94$, and dashed: $\rho_0=1.4$. Inset: cortex width as a function of $\rho_0$ 
   from numerical solution of Eqs.~(\ref{eq:stress1d}) (dots) and from Eq.~(\ref{eq:L}) (line). Parameters are 
   $a=-4.5 k_d\eta$ and $b=3.5 k_d\eta$.}
 \label{fig:highactivity2}
\end{figure}

We can calculate the density profile for $\rho\approx\rho_1$. Then $f(\rho)=-\alpha(\rho-\rho_1)$ with $\alpha=|f'(\rho_1)|$ such that
\begin{equation}
\int_{\rho_0}^\rho \frac{k_d\eta}{\rho'f(\rho')}d\rho'\approx -\frac{k_d\eta}{\alpha\rho_1}\ln\frac{|\rho-\rho_1|}{|\rho_0-\rho_1|}
\end{equation}
and
\begin{equation}
\frac{d\rho}{dz} =-\frac{\alpha\rho_1^2}{\rho_0\eta v_p}(\rho-\rho_1)\left[\frac{|\rho-\rho_1|}{|\rho_0-\rho_1|}\right]^{-k_d\eta/\alpha\rho_1}
\quad.
\end{equation}
We can solve the latter equation explicitly. In the case $\rho_0\approx\rho_1$ we get for the whole density profile
\begin{equation}
\rho(z) = \rho_1 + (\rho_0-\rho_1)\left\{\frac{\rho_1}{\rho_0}\frac{k_d}{v_p}(L-z)\right\}^{\alpha\rho_1/k_d\eta}.
\label{eq:rhoz}
\end{equation}
Note, that the density profile in the vicinity of $L$ depends on the value of $\alpha\rho_1/k_d\eta$ that fixes the slope of the profile 
when $\rho = \rho_1$. In Figures~\ref{fig:highactivity} and \ref{fig:highactivity2}, we present steady state solutions for various nucleator 
densities $\rho_0$ for $\alpha\rho_1/k_d\eta<1$ and $\alpha\rho_1/k_d\eta>1$, respectively.

From the approximate density profile (\ref{eq:rhoz}), we estimate the cortex thickness $L$ by setting $z=0$, yielding
\begin{equation}
L\sim\frac{\rho_0}{\rho_1}\frac{v_p}{k_d}\quad.
\label{eq:L}
\end{equation}
As can be seen in Figs.~\ref{fig:highactivity} and \ref{fig:highactivity2}, for $\rho\approx\rho_0$ it approximates well the thickness 
obtained from the numerical 
solution to Eqs.~(\ref{eq:stress1d}). For $\rho_0>\rho_1$ the length increases linearly with $\rho_0$, however, with a slope that
is different from the one given in Eq.~(\ref{eq:L}). 

\section{Discussion}

In this work, we have introduced a hydrodynamic theory for describing the dynamics of active gels in presence of
filament polymerization and depolymerization. Our description of filament assembly and disassembly requires 
some comments. Specifically, we have taken the depolymerization rate and
viscosity to be constant. Our main
results are more general and do not depend on this assumption. The same behavior is obtained if we keep the gel 
density dependences of depolymerization rate and viscosity. In general, however, the depolymerization 
process is more complex as the rate of filament disassembly depends on mechanical
stresses, the degree of severing, and the presence of actin associated proteins like cofilin or gel-solin. In addition the fact 
that filaments depolymerize from their ends might introduce a gradient in the effective depolymerization rate. Together these
processes might even lead to filament subpopulations with different turnover rates~\cite{fritz13}. Similarly, the polymerization process
is more involved than assumed in the present work. We have restricted it to be 
spatially localized at the membrane, while actin filaments
also grow in the bulk and can be nucleated away from the membrane. This possibility will be discussed in a forthcoming 
publication. We believe, however, that linking cortex formation to an out of equilibrium wetting phenomenon opens a new way of thinking about the problem
worth being fully investigated. This picture is a non-equilibrium 
analog of a (pre-)wetting transition whereby the actin condensation on the membrane is driven by the contractility of the myosin
molecular motors. For sufficiently strong motor activity, the actin density is almost constant up to a certain thickness, where
it drops sharply to zero. Our simplified theory produces a singularity at the edge of the layer. This singularity can be smoothed out by adding 
terms in the osmotic pressure depending on the gradient of the density as is classically done in Ginzburg-Landau theory. The 
thickness of the non-equilibrium wetting layer is determined by the polymerization velocity at the membrane and the depolymerization
rate. 

The description of the cortical actin layer as an active wetting layer suggests
several extensions of this work. The interaction between two adjacent cortical
layers is obviously important in situations when the cell thickness becomes
small. This is
notably the case for the lamellipodia of cells crawling on a solid substrate, for example, keratocyte cells. Under these conditions
one might expect a non-equilibrium analog of capillary condensation~\cite{rowl89,rest00}. While in this
work we have focused on steady state properties of the cortical layer, we have also started to study the dynamics of cortex formation
as well as dynamic instabilities of the cortex. In addition, our theory can be applied to in vitro experiments on cell extracts that
contract in presence of actin assembly and disassembly~\cite{bend08}. 

\begin{acknowledgement}
We thank M. Fritzsche and G. Charras (University College London) for kindly providing Fig.~1a, D. Cuvelier (Institut Curie) for 
sharing unpublished results
that have motivated our work, and E. Paluch (MPI for Molecular Cell Biology and
Genetics), Abhik Basu (SINP Kolkata) and Ananyo Maitra (IISc Bangalore) for
stimulating discussions.
J.-F.J. and K.K. thank the German-French University (DFH-UFA) for
financial support through grant G2RFA-133-07. J.-F.J. and J.P. also acknowledge support from the European Network MITOSYS and
J.-F.J., J.P. and S.R.  support from CEFIPRA grant 3504-2, and SR a J C
Bose Fellowship from the DST, India.
\end{acknowledgement}

\end{document}